\documentclass[12pt, a4paper]{article}

\usepackage[utf8]{inputenc}
\usepackage{csquotes} 
\usepackage{graphicx}
\usepackage{caption}
\usepackage{amsmath}
\usepackage[sorting=none]{biblatex} 
\usepackage{geometry}
\usepackage{siunitx} 
\usepackage{float}
\usepackage{hyperref}
\usepackage[table,xcdraw]{xcolor} 
\usepackage{gensymb} 

\graphicspath{{images/}}

\begin{titlepage}
\title{\textbf{Quantum chemical study of the influence of torsional deformation on the properties of chiral WXY (X, Y = S, Se) Janus-nanotubes}}

\vspace{10cm}

\author{Ilya Mikhailov, Anton Domnin, Robert Evarestov \\ Saint-Petersburg State University}

\date{August 2024}

\end{titlepage}

\begin{document}

\maketitle
\thispagestyle{empty}


\begin{abstract}
\normalsize
    This work sheds light on the electronic band properties of chiral WSSe Janus nanotubes from the quantum mechanical perspective. Line groups theory developed by Damnjanovich was used to model mechanical torsion of chiral nanotubes with different symmetries. Known natural torsion in chiral nanotubes was observed. It was shown that mechanical stress can be used as a tool to tune electronic properties of the nanotubes including the change of the nature of the electron transition. On the DFT-level of the theory the stability of WS$_{\text{2}}$, WSe$_{\text{2}}$ and Janus WSSe nanotubes was estimated. Applicability of stable Janus-nanotubes to photocatalytic water-splitting was suggested based on the calculated electronic properties.
\end{abstract}

{\let\clearpage\relax \tableofcontents} 

\pagebreak

\section{Introduction}

Janus nanomaterials have long been known to researchers causing exclusive interest. In 2024 WSSe-Janus monolayer was first synthesized via CVD-substitution reaction of WSe$_2$ monolayer with highly reactive hydrogen plasma passing through a sulfur substrate \cite{kaneda}. Remarkably, it was also shown that these asymmetric monolayers were subsequently rolled into nanoscrolls, which suggests the possibility of new approach to the  synthesis of Janus WSSe-nanotubes.

Asymmetric layer's tendency to curl was foreseen by Linus Pauling long before the nanotubes were synthesized \cite{pauling1930structure}. He suggested that driving force for curling originated from the difference in the lattice parameters of inner and outer layers of the material. Properties of Janus-nanotubes discussed in this study are very different from their parental WS$_\text{2}$ and WSe$_\text{2}$ structures \cite{jucat}. In this work, energies and electronic properties of WS$_\text{2}$, WSe$_\text{2}$, WSSe$_\text{in}$ and WSSe$_\text{out}$ nanotubes were compared using quantum chemistry methods.

The \textit{ab initio} study of the effect of the mechanical deformation (twisting and axial strain) on transition metal dichalcogenide's nanotubes (including their Janus versions) properties is present in literature \cite{bhard2022strain, bhard2024strain}. However, there are several reasons for this study to appear. First, few works are devoted to chiral nanotubes, although they are the ones most often observed in the experiment \cite{an2024direct}. Theorists favor zigzag and armchair nanotubes which have higher symmetry and are easier to simulate. Second, results of the modeling of electronic properties of WSSe-Janus nanotubes under torsion deformations can barely be found in literature. Thus, the aim of this work is to deliver the insights of the \textit{ab initio} study of chiral WSSe-Janus nanotubes under torsion deformations.

\section{Modeling methodology}

Since the methodology used in this work is based on the line group theory, a brief discussion of relevant basic principles is given below.

\subsection{Introduction to Line Groups}

Nanotubes are quasi-1D structures and their symmetry is described with line symmetry groups, a theory which was developed by Damnjanovich \cite{damnj}. He introduced the following factorization of the line group: $\textbf{\textit{L}}=\textbf{\textit{Z}}\cdot\textbf{\textit{P}}$ where \textbf{\textit{Z}} is a \textit{generalized translation} group and \textbf{\textit{P}} is the point symmetry group of a monomer. Unlike achiral WX$_\text{2}$ nanotubes (where X is chalcogene atom), chiral nanotubes possess only helical and rotation $C_{n}$ symmetry. The subgroup \textbf{\textit{Z}} only consists of operations $(C_Q|f)$ which are rotations around the screw axis of order $Q$ followed by translations along the nanotube's axis to $f$. When the $Q$ is rational number, the structure is called \textit{commensurate} and we can write $Q = \frac{q}{r}$ where $q$ and $r$ are co-primes. Otherwise, the structure is \textit{incommensurate} and lacks translational periodicity. The line symmetry group of WX$_2$ chiral nanotube is denoted as $\textbf{\textit{L}} = \textbf{\textit{T}}_{q}^{r}\cdot\textbf{\textit{C}}_n$. This group belongs to the first family of 13 line groups families. This classification is based on the fact that symmetry operations preserve the direction of the axis up to its sign. Achiral WX$_2$ nanotubes have point symmetries $\textbf{\textit{P}} = \textbf{\textit{C}}_{nh}$ (armchair) and $\textbf{\textit{P}} = \textbf{\textit{C}}_{nv}$ (zigzag) belonging to 4th and 8th families respectively. The characteristic angle of the nanotubes is determined by $Q$: $\frac{2\pi}{Q}=\varphi = \frac{2\pi r}{q}$. The torsion angle, a measure of the twisting, can be determined by $\omega = \varphi_0 - \varphi_x$ where $\varphi_0$ is the angle of pristine nanotube. Surely, the distorted structure has a new symmetry that corresponds to $\varphi_x = \frac{2\pi}{Q_x}$.

To obtain crystallographic notation $\textbf{\textit{L}}q_p$ of the symmetry group of WX$_2$ nanotube, the displacement index $p$ must be determined. It can be done via formula 1 \cite{damnj}:

\begin{equation}
rp = lq + n \quad\quad l = 0, 1, 2...
\label{formula}
\end{equation}

Where $n$ is the greatest common divisor (GCD) of the nanotube's chirality indices $(n_1, n_2)$. The crystallographic notation helps to define the structure of the nanotube for a simulation software. Thus, numerical search allows one to generate several distorted nanotubes in a given range of $\omega$.

In this work nanotubes with chirality indices $(8, 2)$ and $(12, 3)$ were studied. Chirality indices $(n_1, n_2)$ are sufficient to determine the undistorted (pristine) nanotube's symmetry \cite{domnin2023dft}. These nanotubes lie on the one translation vector and have line symmetry groups $\textbf{\textit{L}}28_{18}$ and $\textbf{\textit{L}}42_{27}$ respectively (in crystallographic notation). The $\textbf{\textit{L}}=\textbf{\textit{Z}}\cdot\textbf{\textit{P}}$ factorization is then $\textbf{\textit{T}}_{28}^{11}\cdot\textbf{\textit{C}}_2$ and $\textbf{\textit{T}}_{42}^{11}\cdot\textbf{\textit{C}}_3$ for $(8, 2)$ and $(12, 3)$ chiralities. All of the symmetries for these nanotubes and their distorted versions are listed in table 1.

\begin{table}[H]
\centering
\caption{Torsion angles of the nanotubes under study and their symmetry groups}
\begin{tabular}{c|c|c}
\rowcolor[HTML]{EFEFEF} 
{\color[HTML]{333333} $\omega$,$\degree$} & {\color[HTML]{333333} (8, 2)} & {\color[HTML]{333333} (12, 3)} \\ \hline
-2.857                      & $\textbf{\textit{L}}18_8$ ($\textbf{\textit{T}}_{18}^{7}\textbf{\textit{C}}_2$)                         & $\textbf{\textit{L}}27_{12}$ ($\textbf{\textit{T}}_{27}^{7}\textbf{\textit{C}}_3$)                        \\
-1.607                      & $\textbf{\textit{L}}64_{18}$  ($\textbf{\textit{T}}_{64}^{25}\textbf{\textit{C}}_2$)                      & $\textbf{\textit{L}}96_{27}$ ($\textbf{\textit{T}}_{96}^{25}\textbf{\textit{C}}_3$)                         \\
-1.118                      & $\textbf{\textit{L}}46_{18}$ ($\textbf{\textit{T}}_{46}^{18}\textbf{\textit{C}}_2$)                        & $\textbf{\textit{L}}69_{27}$ ($\textbf{\textit{T}}_{69}^{18}\textbf{\textit{C}}_3$)                        \\
-0.695                      & $\textbf{\textit{L}}74_{46}$ ($\textbf{\textit{T}}_{74}^{29}\textbf{\textit{C}}_2$)                        & $\textbf{\textit{L}}111_{69}$ ($\textbf{\textit{T}}_{111}^{29}\textbf{\textit{C}}_3$)                       \\
-0.326                      & $\textbf{\textit{L}}158_{130}$ ($\textbf{\textit{T}}_{158}^{62}\textbf{\textit{C}}_2$)                     & $\textbf{\textit{L}}237_{195}$ ($\textbf{\textit{T}}_{237}^{62}\textbf{\textit{C}}_3$)                      \\
0.000                       & $\textbf{\textit{L}}28_{18}$ ($\textbf{\textit{T}}_{28}^{11}\textbf{\textit{C}}_2$)                        & $\textbf{\textit{L}}42_{27}$ ($\textbf{\textit{T}}_{42}^{11}\textbf{\textit{C}}_3$)                        \\
0.343                       & $\textbf{\textit{L}}150_{28}$ ($\textbf{\textit{T}}_{150}^{59}\textbf{\textit{C}}_2$)                      & $\textbf{\textit{L}}225_{42}$ ($\textbf{\textit{T}}_{225}^{59}\textbf{\textit{C}}_3$)                       \\
0.547                       & $\textbf{\textit{L}}94_{28}$ ($\textbf{\textit{T}}_{94}^{37}\textbf{\textit{C}}_2$)                       & $\textbf{\textit{L}}141_{42}$ ($\textbf{\textit{T}}_{141}^{37}\textbf{\textit{C}}_3$)                       \\
1.353                       & $\textbf{\textit{L}}38_{28}$ ($\textbf{\textit{T}}_{38}^{15}\textbf{\textit{C}}_2$)                       & $\textbf{\textit{L}}57_{42}$ ($\textbf{\textit{T}}_{57}^{15}\textbf{\textit{C}}_3$)                        \\
2.143                       & $\textbf{\textit{L}}48_{38}$ ($\textbf{\textit{T}}_{48}^{19}\textbf{\textit{C}}_2$)                       & $\textbf{\textit{L}}72_{57}$ ($\textbf{\textit{T}}_{72}^{19}\textbf{\textit{C}}_3$)                        \\
3.025                       & $\textbf{\textit{L}}68_{58}$ ($\textbf{\textit{T}}_{68}^{27}\textbf{\textit{C}}_2$)                       & $\textbf{\textit{L}}102_{87}$ ($\textbf{\textit{T}}_{102}^{27}\textbf{\textit{C}}_3$)                      
\end{tabular}
\end{table}

As can be seen from table 1, the nanotube with the largest unit cell has symmetry $\textbf{\textit{L}}237_{195}$ with $237 \times 3 = 711$ atoms in the unit cell (237 W atoms and 474 atoms of chalcogene). Without imposing symmetry, it would be burdensome to simulate this structure.

\subsection{Computational Details}

Quantum mechanical simulations were carried out using CRYSTAL17 software \cite{dovesi1, dovesi2}. An outstanding feature of this software is the possibility to set an arbitrary symmetry group for subperiodic structures (e.g. line groups for nanotubes) in matrix format. Hybrid HSE06 functional with 25\% of HF exchange was exploited with CRENBL basis set and D2 dispersion correction. This basis set includes effective core potential ECP for all atoms to take relativistic effects into account \cite{ross1990ab}. Convergence criteria for energy self-consistence was set to $10^{-9}$ a.u. Strict accuracy criteria $10^{-8}, 10^{-8}, 10^{-8}, 10^{-8}, 10^{-16}$ a.u. for Coulumb and exchange integrals calculation was imposed. The reciprocal lattice integration was performed by sampling Brillouin zone with 18x1x1 Monkhorst-Pack mesh \cite{monkhorst1976special} resulting in 10 special points. Validation of this scheme was provided in our previous work \cite{domnin2023dft}.

\section{Results and Discussion}

The diameter of the hexagonal nanotube can be found via formula 2:
\begin{equation}
    D = \dfrac{a}{\pi}\sqrt{n_1^2+n_1n_2+n_2^2}
    \label{formula2}
\end{equation}

Where $(n_1, n_2)$ are chirality indices and $a = 3.16$ \r{A} is the monolayer hexagonal lattice parameter. In this study, diameter of the nanotube is counted as doubled distance from symmmetry axis to $W$ atoms. Therefore, for initial structures, the starting diameters are listed in table 2.

\begin{table}[H]
\centering
\caption{Initial diameters of generated nanotubes without deformation}
\begin{tabular}{c|c|c}
     & $D_{(8, 2)}$, \r{A} & $D_{(12, 3)}$, \r{A} \\ \hline
WS$_2$  & 9.22                              & 13.84                              \\
WSe$_2$ & 9.60                              & 14.40                              \\
WSSe & 9.30                              & 13.92                             
\end{tabular}
\end{table}

After the geometry optimization, diameters of \textit{all} nanotubes remained in the intervals of $10.0 < D_{(8,2)} < 11.5$ \r{A} and $14.5 < D_{(12, 3)} < 15.8$ \r{A} indicating that ruptures did not occur. 

\subsection{Energy Torsional Curves of Nanotubes}

Thermodynamic stability of a nanotube with respect to unfolding might be estimated by computing so-called strain energy:
\begin{equation}
E_{str} = \dfrac{E_{NT}}{N_{NT}} - \dfrac{E_{mono}}{N_{mono}}
    \label{formula3}
\end{equation}

Where $N_{NT}$ and $N_{mono}$ is the number of atoms in unit cells of the nanotube and corresponding monolayer. Basically, this is the difference in electronic energies per atom between the nanotube and the layer. Energy curves $E_{str}(\omega)$ are presented on figs 1, 2.\\
\noindent
\begin{minipage}{.5\textwidth}
    \includegraphics[width=\textwidth]{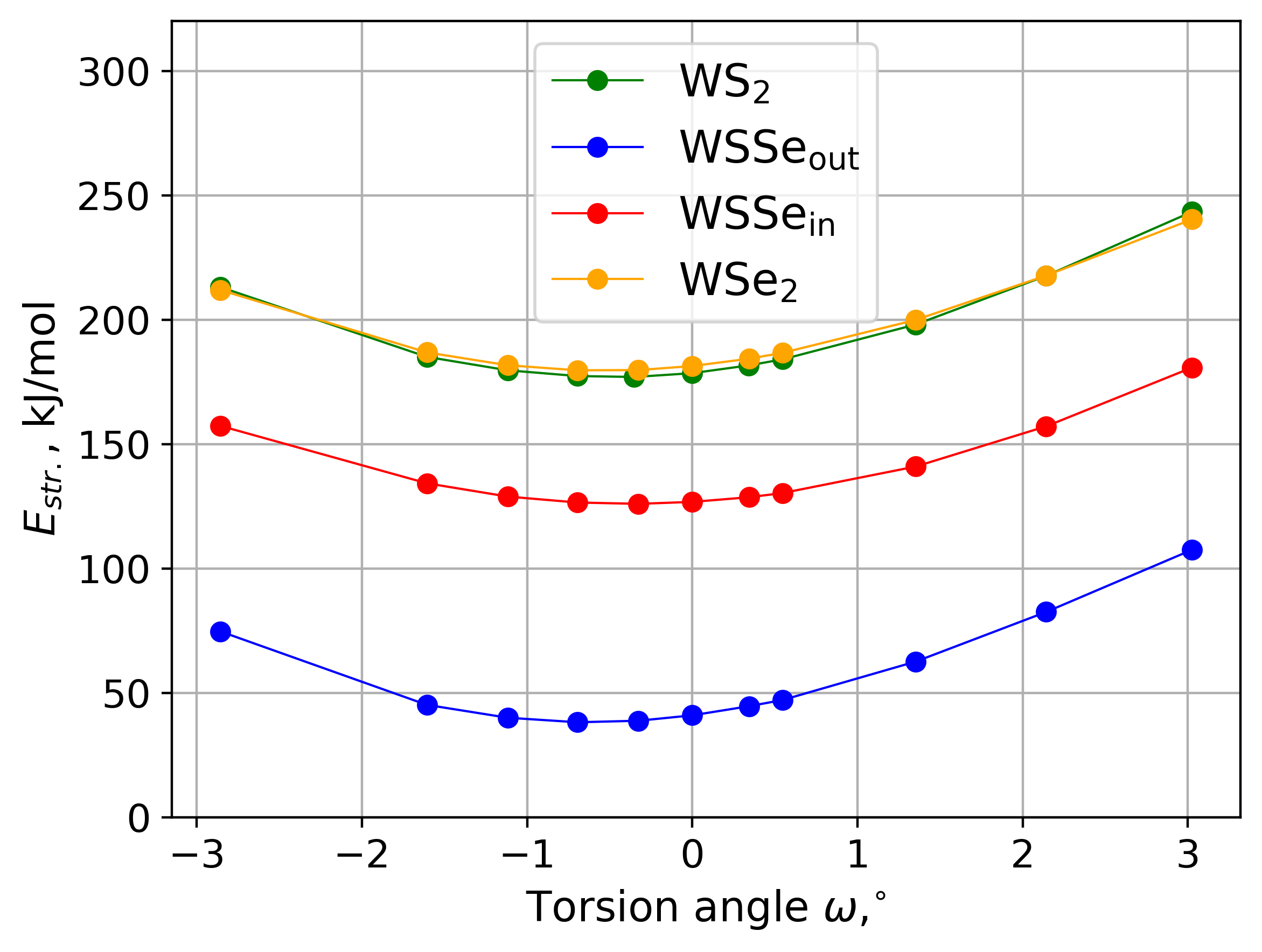}
        \captionof{figure}{Strain energies of (8, 2) NTs}
\end{minipage}%
\begin{minipage}{.5\textwidth}
    \includegraphics[width=\textwidth]{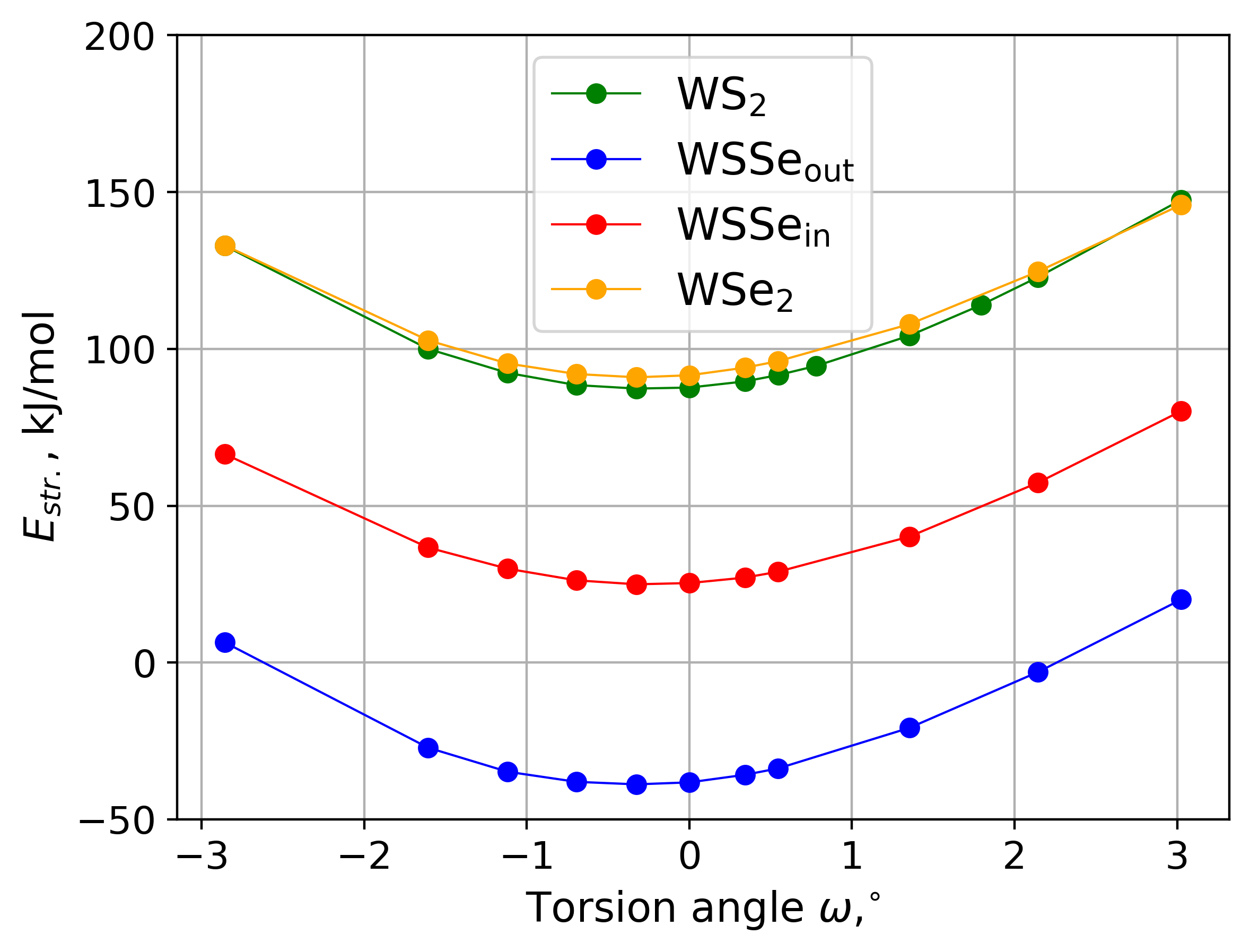}
        \captionof{figure}{Strain energies of (12, 3) NTs}
\end{minipage}
\\
\newpage
It can be seen that strain energies are negative for $(12, 3)$ WSSe$_{\text{out}}$ nanotubes, which is consistent with the experimental observation of the rolling of this Janus monolayer. A careful assumption on the stability of such Janus nanotubes with small diameters of $\approx 15$ \r{A} can be made. \\

It is also clear that for both $(8, 2)$ and $(12, 3)$ nanotubes these energy curves $E(\omega)$ are asymmetric with respect to $\omega = 0 \degree$. To take a closer look on this asymmetry, one can analyze the dependence of the relative energies (equation 4) on the torsion angle (figs 3 and 4).
\begin{equation}
    \Delta E_{rel} = \dfrac{E(\omega)}{N}-\dfrac{E(0)}{N}
    \label{eq4}
\end{equation}
\\
\noindent
\begin{minipage}{.5\textwidth}
    \includegraphics[width=\textwidth]{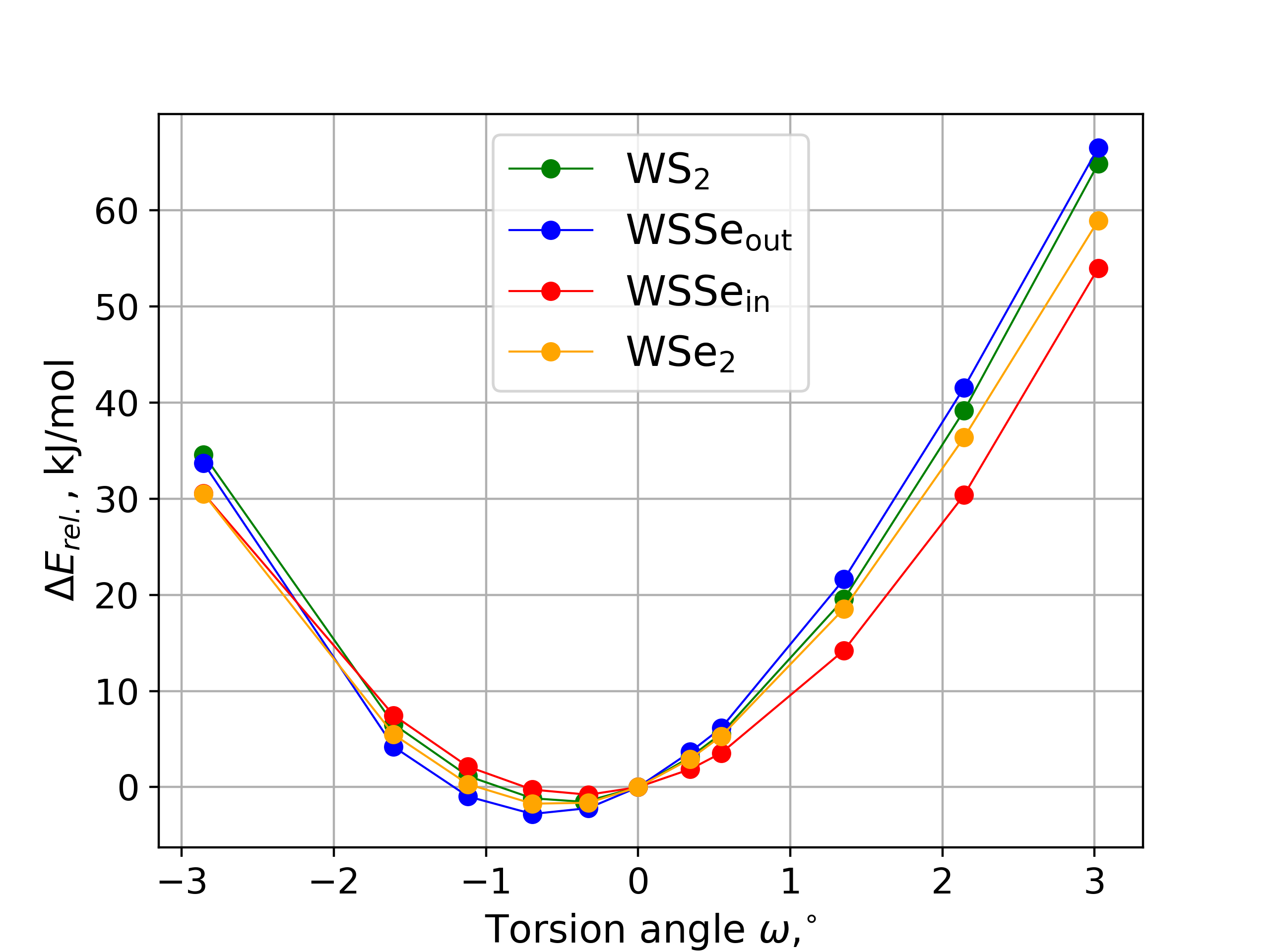}
        \captionof{figure}{Relative energies of (8, 2) NTs}
\end{minipage}%
\begin{minipage}{.5\textwidth}
    \includegraphics[width=\textwidth]{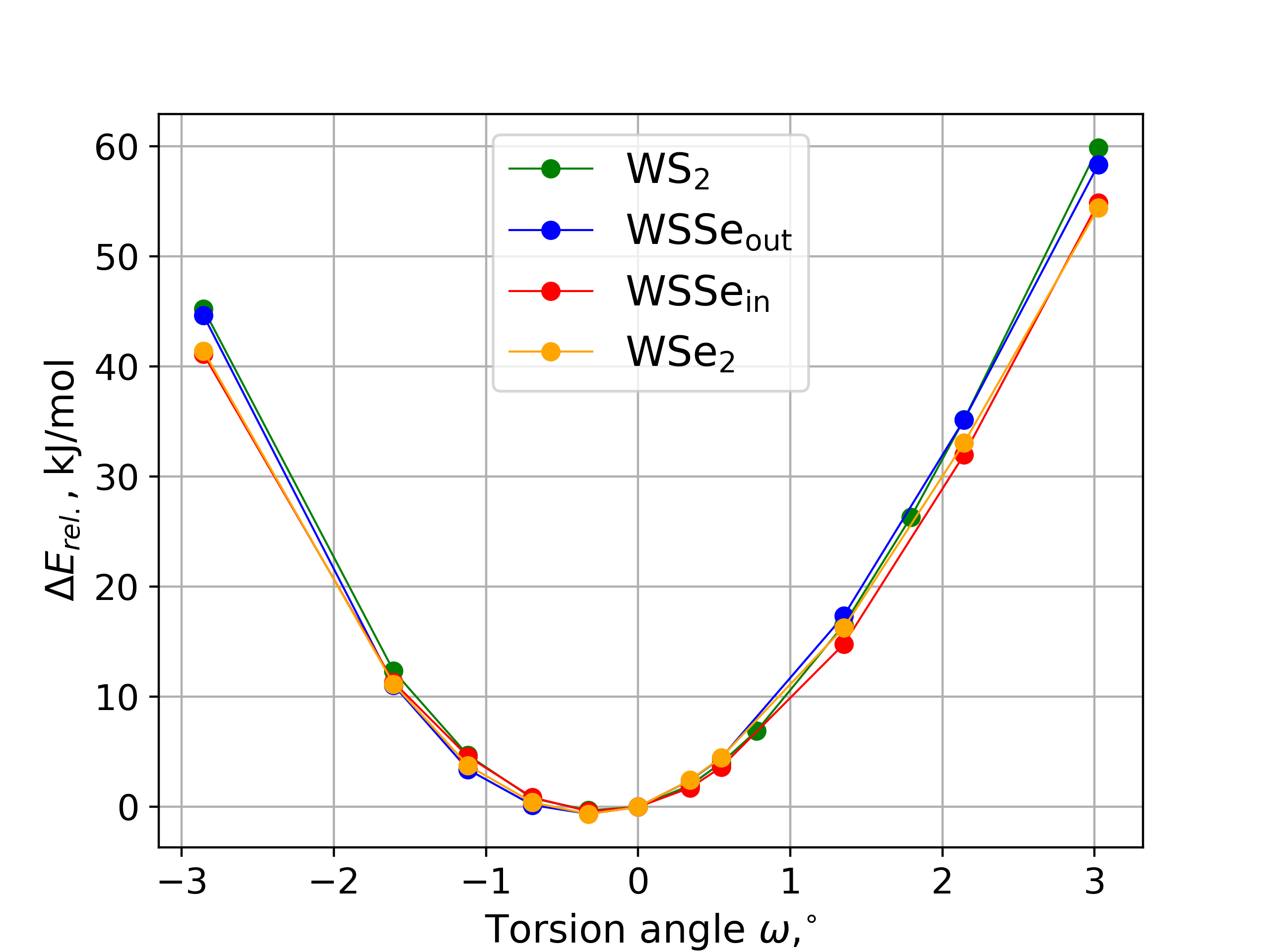}
        \captionof{figure}{Relative energies of (12, 3) NTs}
\end{minipage}
\\

These asymmetries suggest that a nanotube rolled from a layer must undergo spontaneous torsion deformation with changing its symmetry. For instance, the exact energy minima for $(8, 2)$ WSSe$_{\text{out}}$ nanotube is located in the proximity of $\omega = -0.695\degree$. It means that the true structure of this nanotube is \textit{incommensurate} with irrational order of screw axis and its symmetry cannot be represented by crystallographic notation. \\

Incredibly, this phenomenon of the asymmetry of the torsional curve was observed in 2010 \cite{vercosa2010torsional} for chiral $(5, 3)$ carbon nanotubes. Authors noted that only chiral nanotubes have this shift of energy minima from $\omega = 0\degree$; for achiral nanotubes the $E(\omega)$ is perfectly symmetric. They called this \textquote{\textit{natural torsion}}. With the results presented in this work, one can assume that \textquote{natural torsion} is characteristic for all chiral nanotubes regardless of their composition.

\subsection{Electronic properties}
\subsubsection{One-electron properties}
The increased interest in subperiodic transition metal dichalcogenide nanomaterials is due to their potential application in photocatalytic water splitting. This process is prominent for producing hydrogen \textquote{green energy}. Proposed mechanism of the photocatalysis impose some energetical restrictions on the semiconducting catalytic material \cite{kim2019toward}:
\begin{equation}
    1.6 \text{ eV} (\approx780 \text{ nm}) < E_{gap} < 2.8 \text{ eV} (\approx440 \text{ nm})
\end{equation}
\begin{equation}
    E_{VB} < E_{O_2/H_2O} < E_{H_2/H_2O} < E_{CB}
\end{equation}

Where $E_{gap}$ is a band gap of the material, $E_{VB}$ and $E_{CB}$ are valence band top and conduction band bottom, and $E_{O_2/H_2O}$, $E_{H_2/H_2O}$ are redox potentials of water. Moreover, the electron transition between energies at the edges of electronic bands should be direct to prevent the recombination of the electron-hole pair. In this work the dependence $E_{gap}(\omega)$ was analyzed for $(8,2)$ and $(12,3)$ nanotubes (figs 7,8)
\\
\noindent
\begin{minipage}{.5\textwidth}
    \includegraphics[width=\textwidth]{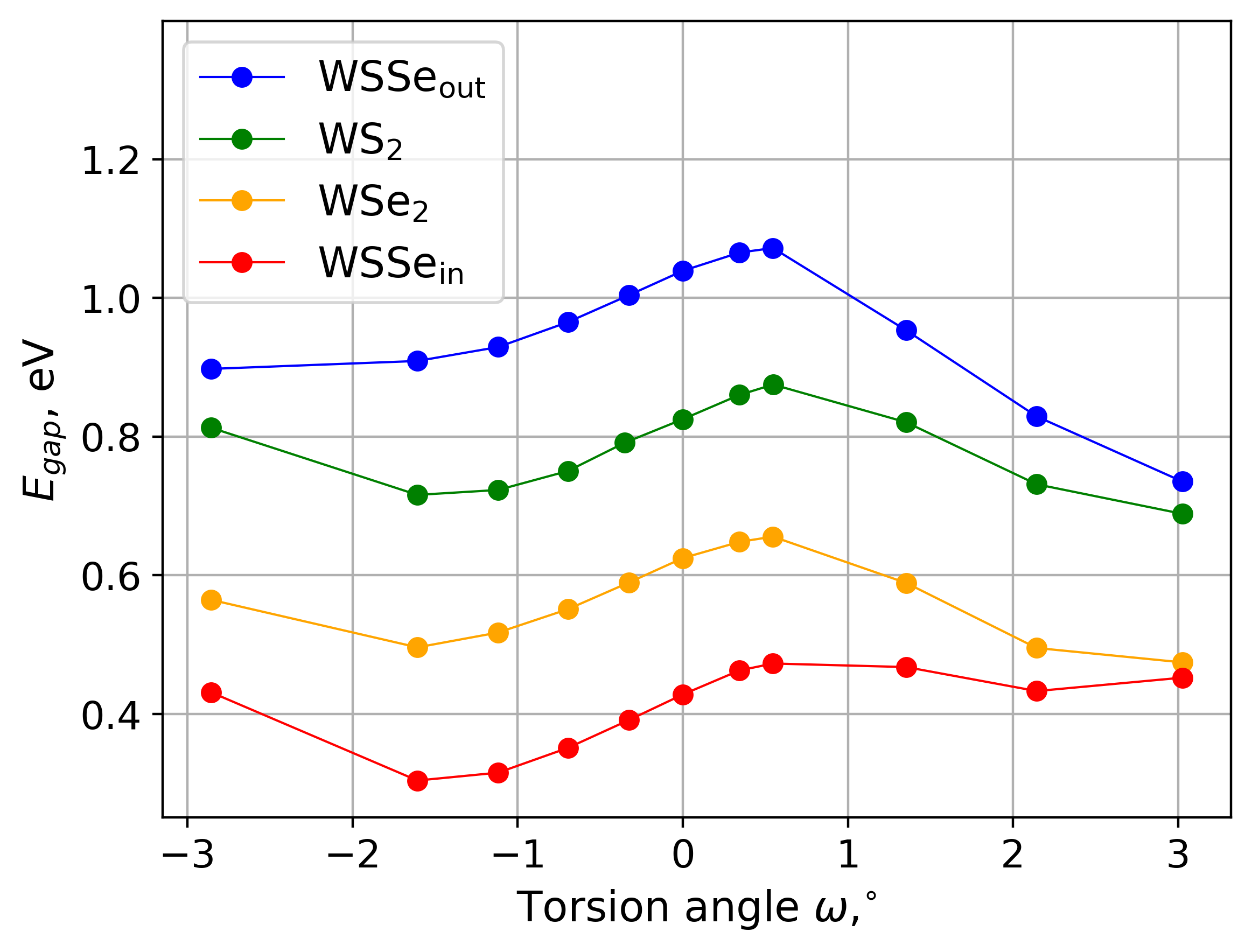}
        \captionof{figure}{$E_{gap}(\omega)$ of (8, 2) NTs}
\end{minipage}%
\begin{minipage}{.5\textwidth}
    \includegraphics[width=\textwidth]{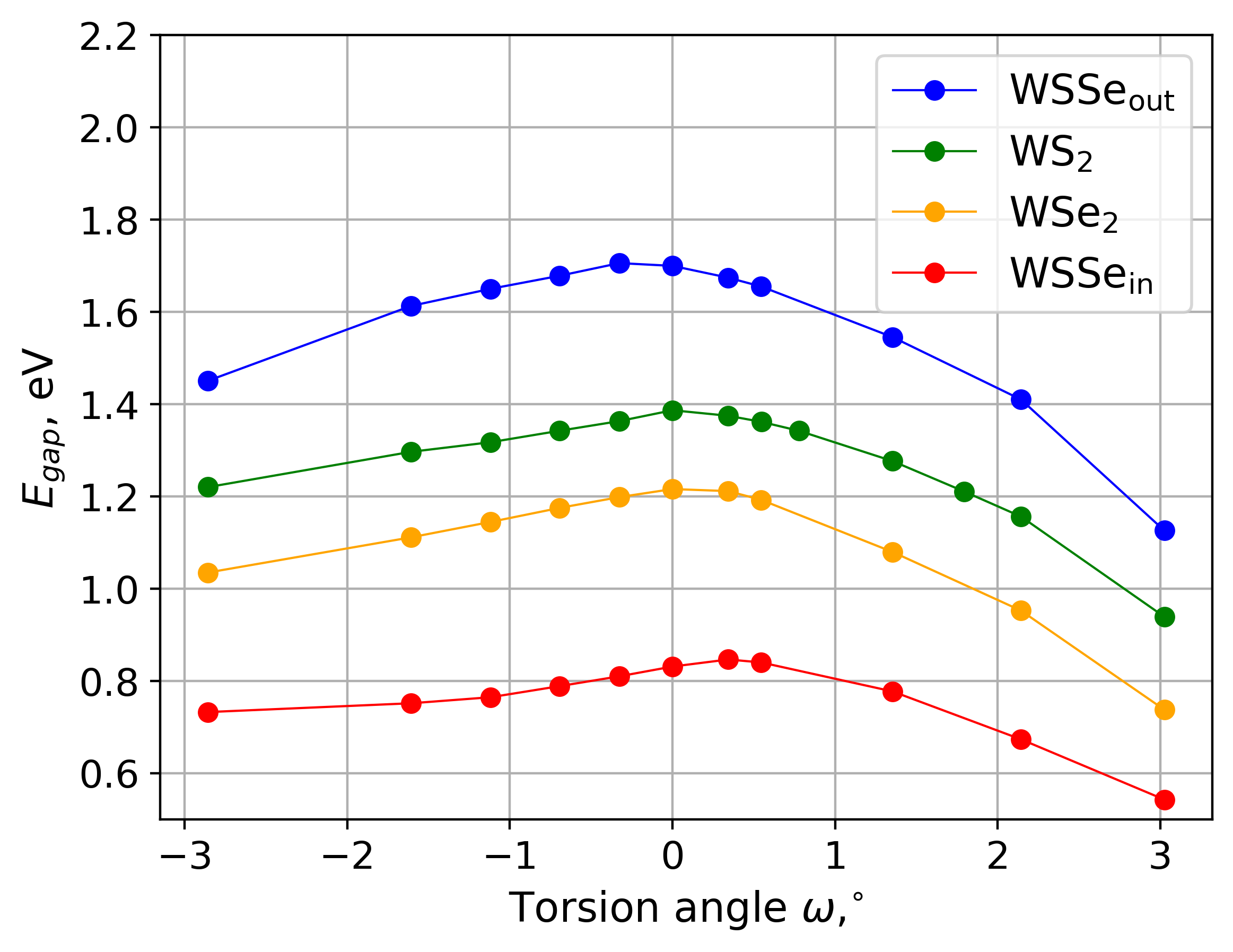}
        \captionof{figure}{$E_{gap}(\omega)$ of (12, 3) NTs}
\end{minipage}
\\

These curves are also asymmetric with respect to $\omega = 0\degree$ because of chiral nature of nanotubes. For the Janus-WSSe$_{out}$ nanotube with chirality $(12, 3)$ the band gap is $E_{gap} > 1.4$ eV in the range $-3.0\degree < \omega < 2.1 \degree$. Thus, the requirement (5) is met. In general, with the increase of torsion, conductivity of a nanotube rises. This effect was observed in the experimental work \cite{ben2022self}. Authors imposed torsional strain on the WS$_2$ nanotube using atomic force microscope needle. \\

Validation of the condition (6) is also important and can be made with figs 9, 10. There, $E_{VB}(\omega)$ and $E_{CB}(\omega)$ are presented for all simulated nanotubes. Dashed black lines correspond to $E_{O_2/H_2O}=-5.66$ eV and $E_{H_2/H_2O} = -4.44$ eV which are standard redox potentials of water.
\\
\noindent
\begin{minipage}{.5\textwidth}
    \includegraphics[width=\textwidth]{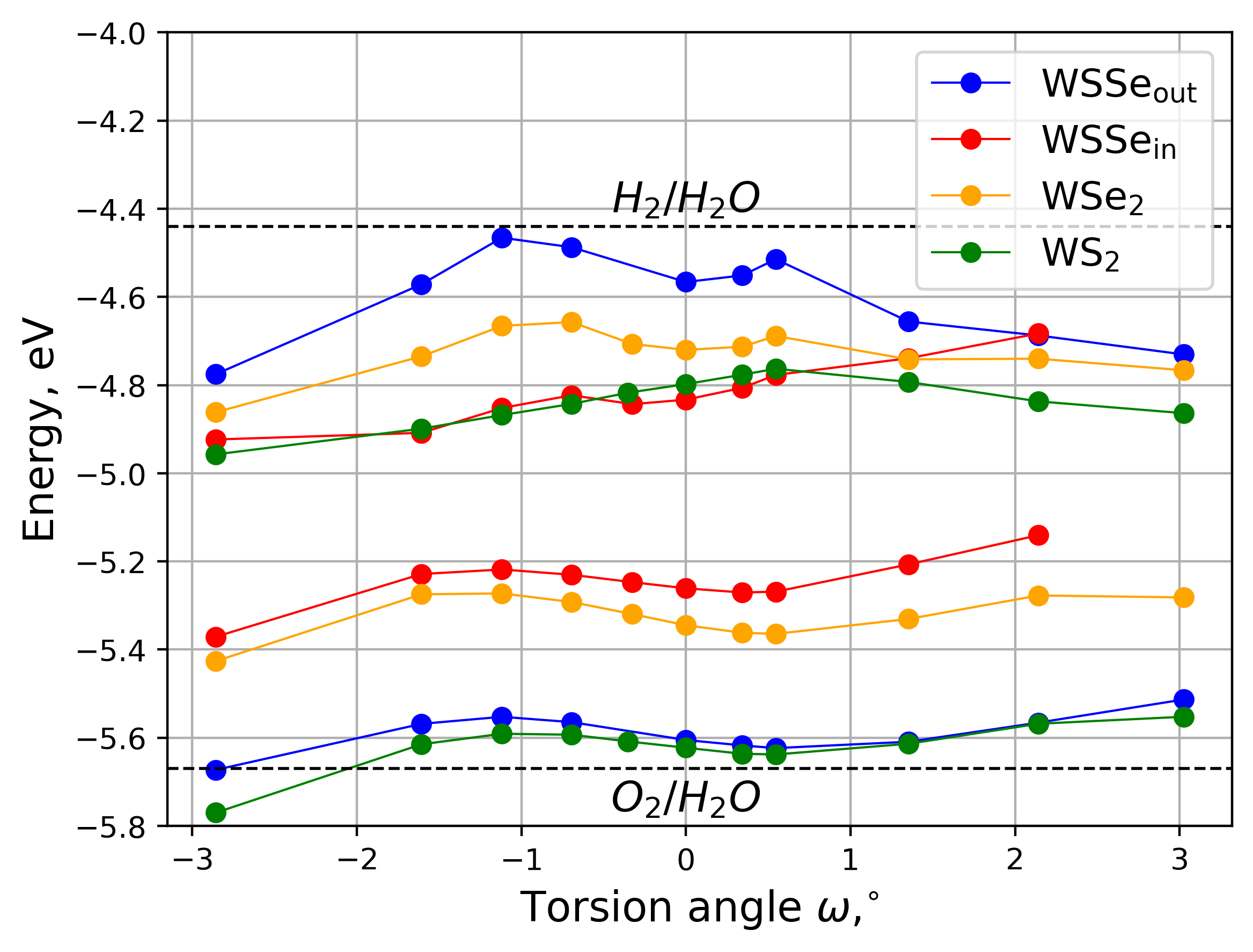}
        \captionof{figure}{$E_{gap}(\omega)$ of (8, 2) NTs}
\end{minipage}%
\begin{minipage}{.5\textwidth}
    \includegraphics[width=\textwidth]{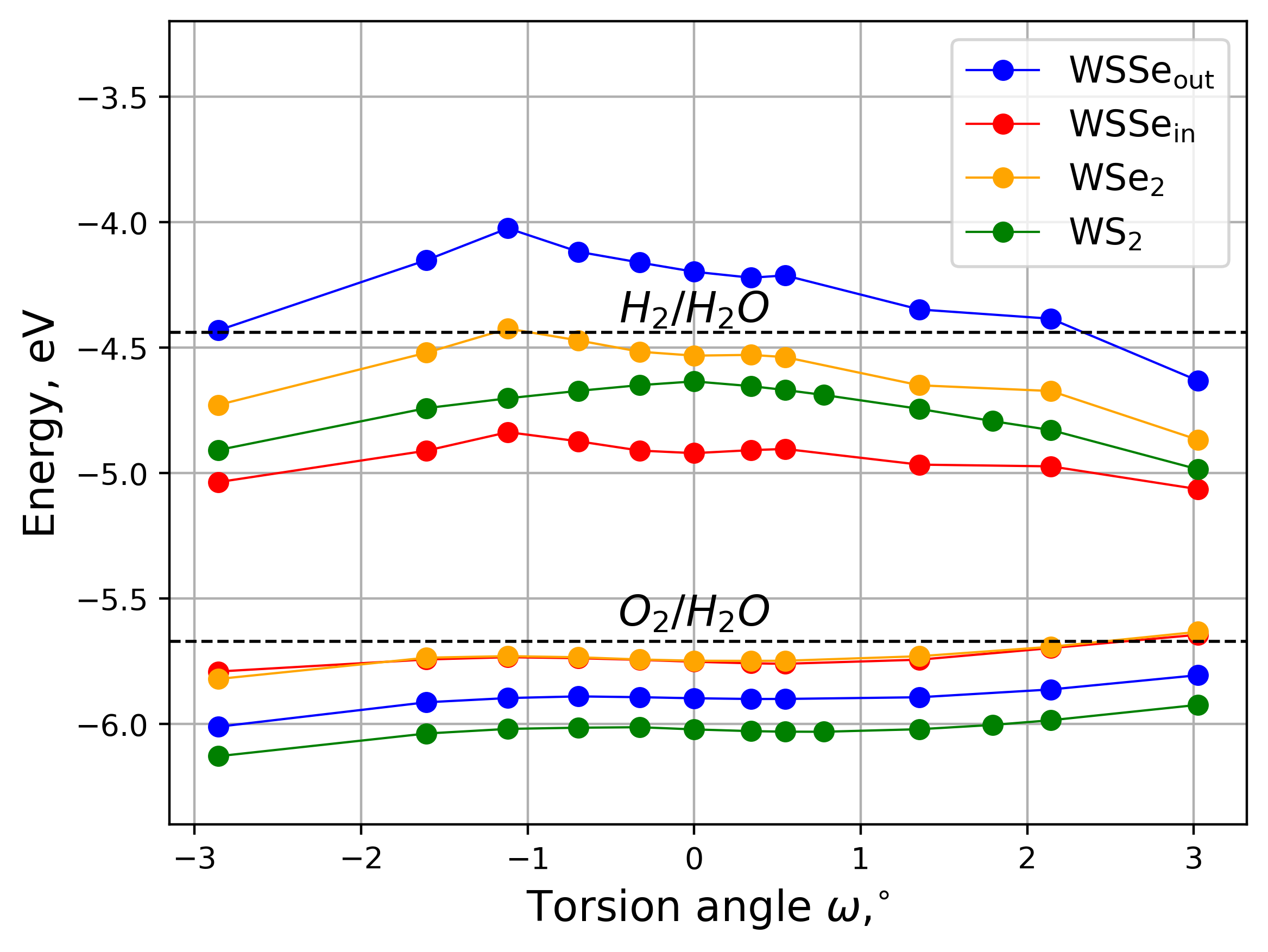}
        \captionof{figure}{$E_{gap}(\omega)$ of (12, 3) NTs}
\end{minipage}
\\

The main contribution to the valence band top is apparently made by the hybridized orbitals of the chalcogene atoms that form the inner wall of the nanotube, since the energy values in the cases of WS$_2$ (green) and WSSe$_{\text{out}}$ (blue) are close to each other over the entire range of the torsion angle; the same is true for the pair WSe$_2$ (orange) – WSSe$_{\text{in}}$ (red). Firstly, regardless of the composition, the top of the valence band is below the oxidation potential of water, i.e. the left side of inequality (6) is satisfied. Secondly, among nanotubes of different compositions, it is Janus WSSe (selenium on the outside) that has a suitable conduction band bottom energy in the range of the torsion angle $-2.8\degree < \omega < 2.2\degree$, thus fully satisfying the condition (6).

\subsubsection{Band structure properties}

To determine the nature of the transition between the top of the valence band and the bottom of the conduction band (direct/indirect), calculations of the electron band structure for several nanotubes were performed. It is appropriate to analyze here the band structures only for $(12, 3)$ WSSe$_{\text{out}}$ nanotube, because it turned out to be stable and satisfactory one-electron properties.

\begin{figure}[H]
\centering
\includegraphics[width=\textwidth]{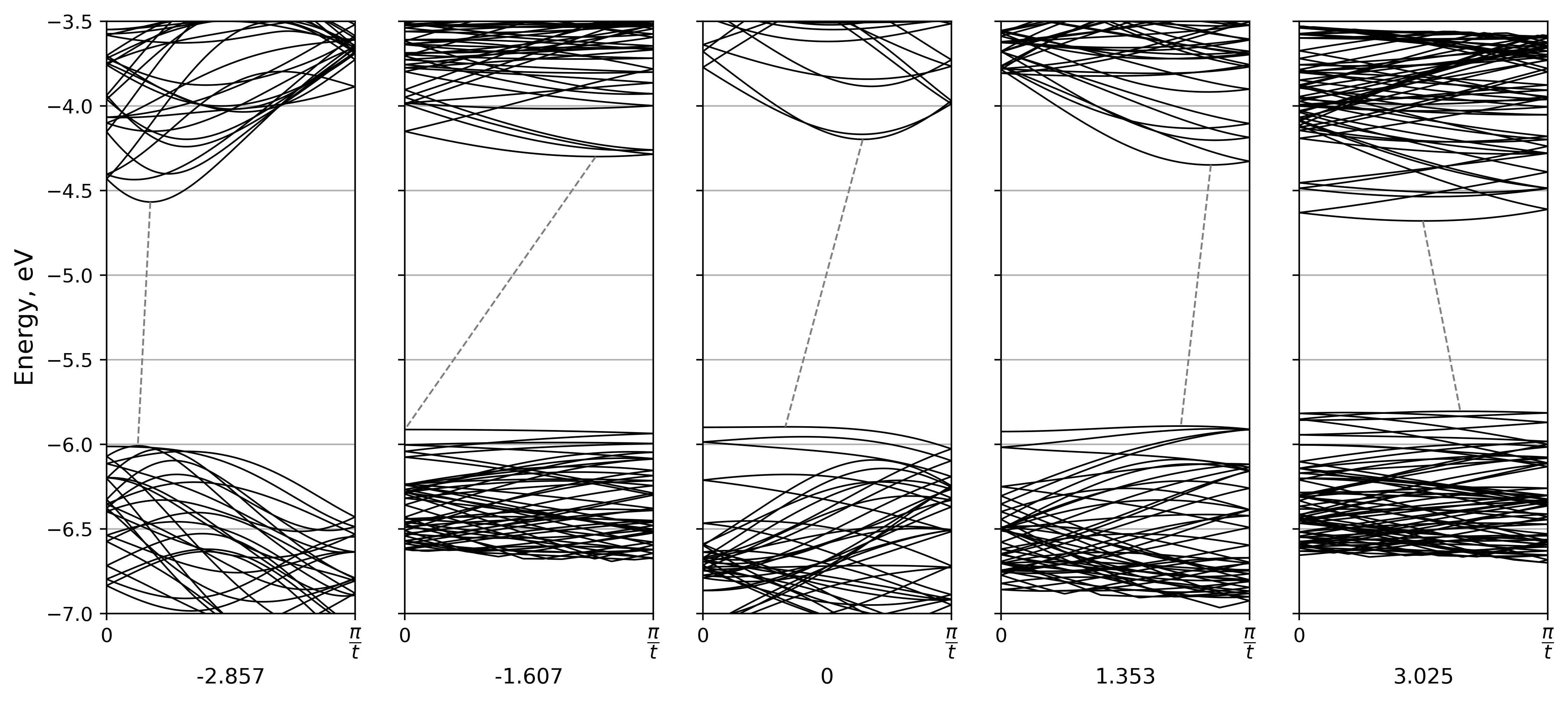}
\caption{Band structures of WSSe$_{\text{out}}$ NTs with $\omega = -2.857\degree$, $-1.607\degree$, $0\degree$, $1.353\degree$, $3.025\degree$}
\label{figure_11}
\end{figure}

Despite the fact that in the majority of cases nanotubes exhibit an indirect type of transition, one can observe the slope of straight line (gray dashed line) connecting the bottom of the conduction band and the top of the valence band. The tangent of the angle of inclination in the mentioned line changes sign when applying torsional deformation. This may serve as an indication that in the angle range of 1.353–3.025\degree the type of transition changes from indirect to direct. The presented bands demonstrate how mechanical deformation influences the electronic properties of the material, which is a significant result for the field of nanomaterials \textquote{tuning}.
\newpage
\section{Conclusions}

It was observed that the energy minima for chiral WX$_2$ nanotubes does not correspond to \textquote{ideal} structure (the nanotube rolled up from a layer). An \textquote{ideal} nanotube must undergo spontaneous torsional deformation with the different symmetry of resulting structure. \\

Janus nanotube WSSe$_{out}$ with chirality $(12, 3)$ and optimized diameter $D \approx 15$\r{A} has negative strain energy and meets formal requirements for photocatalyst in the range of torsion angle $-2.6\degree < \omega < 2.2\degree$. \\

Twisting deformation can be used as a tool to adjust electronic properties of quasi-1D nanostructures. In particular, it was shown that for a certain range of $\omega$ deformation that changes indirect transition to direct can be found.
\newpage
\printbibliography

\end{document}